\documentclass[a4paper, 11pt]{article}
\usepackage{comment} 
\usepackage[nointegrals]{wasysym}
\usepackage{fullpage} 
\usepackage{graphicx}
\usepackage{cite}
\usepackage{bm}
\usepackage{float}
\usepackage{amsmath}
\usepackage[utf8]{inputenc}
\usepackage{amssymb}
\usepackage{mathrsfs}
\usepackage{wrapfig}
\usepackage{enumitem}
\usepackage[english]{babel}
\usepackage{comment}
\usepackage{bbold}  
\usepackage{multicol}
\usepackage[usenames,dvipsnames]{xcolor} 
\usepackage{xcolor,hyperref}
\usepackage{setspace}
\usepackage{abstract}
\usepackage{sectsty}
\sectionfont{\large}
\usepackage{tablefootnote}

\usepackage[T1]{fontenc}
\usepackage[margin=0.94in]{geometry}
\def\Mpl{M_{\rm P}}

\begin{document}

\null\hfill 
IPMU26-0007, KEK-TH-2810, KEK-Cosmo-0412 \\

\vspace*{\fill}
\begin{center}
    \Large\textbf{{ \textcolor{Black}{On the Kalb-Ramond field with non-minimal coupling to gravity} }}\\
    \normalsize\textsc{Anamaria Hell}$^{1,2}$ and \textsc{Ippei Obata}$^{3,1}$ 
\end{center}

\begin{center}
     $^{1}$ \textit{Kavli IPMU (WPI), UTIAS,\\ The University of Tokyo,\\ Kashiwa, Chiba 277-8583, Japan}\\
    $^{2}$ \textit{Center for Data-Driven Discovery, Kavli IPMU (WPI), UTIAS,\\ The University of Tokyo, Kashiwa, Chiba 277-8583, Japan}\\
    $^{3}$ \textit{Theory Center, Institute of Particle and Nuclear Studies (IPNS), High Energy Accelerator Research Organization (KEK), 1-1 Oho, Tsukuba, Ibaraki 305-0801, Japan}
    
\end{center}
\thispagestyle{empty} 

\renewcommand{\abstractname}{\textsc{\textcolor{Black}{Abstract}}}

\begin{abstract}
We consider a massive Kalb-Ramond field with a general non-minimal coupling to gravity. We first study the theory in flat space-time, taking into account the non-linearities. We show that the coupling with the Ricci scalar gives rise to the strong coupling of the two transverse pseudo-vector degrees of freedom, which are absent in the massless theory. We then show that if the theory is instead coupled to the Ricci tensor or the Riemann tensor, the two tensor modes become strongly coupled in addition to the transverse pseudo-vector modes. We then extend our analysis to homogeneous and isotropic space-time, with vanishing background value of the Kalb-Ramond field. We show that in this case, the couplings with the Ricci and Riemann tensor give rise to the runaway instability. Finally,  we discuss the inclusion of the disformal coupling as a possible resolution to this unnatural behavior. 
  \end{abstract}
 
\vfill

\small
\href{mailto:anamaria.hell@ipmu.jp}{{anamaria.hell@ipmu.jp}}\\
\href{mailto:iobata@post.kek.jp}{iobata@post.kek.jp}
\vspace*{\fill}

\clearpage
\pagenumbering{arabic} 
\newpage

\tableofcontents
\newpage

\section{Introduction}

 The presence of strong coupling, the regime in which the perturbation theory breaks down, has become a natural occurrence in various field theories and theories of gravity. While this effect is often referred to as a pathology of the theory, strictly speaking, it indicates a shortcoming of standard approaches, and a need for a different set of methods or tools to infer the non-perturbative nature of a theory. 

One can expect that strong coupling occurs whenever one introduces a new degrees of freedom by modifying a theory, and then takes the limit in which this modification vanishes \cite{ Hell:2025pso}. A well known example of this is the scalar mode, which appears in addition to the two vector modes when graviton is made massive. Due to it, at the linearized order, the theory of a massive graviton with a Fierz-Pauli  mass term gives rise to the vDVZ discontinuity -- the discrepancy for the precession of the perihelia of Mercury or the deflection of starlight when massive gravity is compared to the linearized general relativity \cite{Fierz:1939ix, vanDam:1970vg, Zakharov:1970cc, Iwasaki:1970hrg}. However, at the Vainshtein radius — the scale at which the non-linear terms become of the same order as linear ones — the scalar mode becomes strongly coupled, and decouples from the remaining dof up to small corrections\footnote{The vector modes can also be expected to become strongly coupled, as was shown in massive mimetic gravity \cite{Chamseddine:2018gqh}. } \cite{Vainshtein:1972sx, Deffayet:2001uk, Gruzinov:2001hp}. Thus, the strong coupling and decoupling of the scalar mode for scales beyond the Vainshtein radius restores agreement between massive and massless gravity up to small corrections. Another example is massive Yang-Mills theory, in which the mass is added \textit{by hand}. This modification introduces longitudinal (scalar) modes in the theory, which give rise to the singular behaviour of the perturbative series. However, due to the non-linear terms, these modes become strongly coupled at the strong coupling scale, beyond which they decouple from the remaining vector degrees of freedom up to small corrections that become smaller as one approaches higher energies \cite{Hell:2021oea, Hell:2025pso}. Notably, the transverse modes that remain also in the massless theory are weakly coupled both below and beyond the strong coupling scale. This means that the massless limit of massive Yang-Mills theory is smooth, as originally conjectured in \cite{Vainshtein:1971ip}, and shown in \cite{Hell:2021oea}.  In other words, the presence of strong coupling does not indicate that the perturbation theory breaks down for all of the modes, but only those that are absent in the massless theory (or the one prior to a modification that introduces these additional modes).

What is common with massive gravity and massive Yang-Mills theory, as well as many other theories with strong-coupling (see eg. \cite{Hell:2021wzm, Hell:2023mph}) is that the modes that are present in both massless and masive theories are weakly coupled for all scales. The modes that appear by modifying the theories become strongly coupled, but they affect the original ones only up to small corrections. This is also natural — a theory of a photon should not drastically change if its mass is zero, or $\mathcal{O}(10^{-18})$ eV, its current experimental bound \cite{Goldhaber:2008xy,ParticleDataGroup:2024cfk}. However, recently, it was pointed out in \cite{Hell:2024xbv} that if one allows it to interact with gravity through non-minimal coupling, it can give rise to an unexpected behavior of the gravitational waves: If one couples a massive vector field to the Ricci tensor, then no matter how small its mass is, the gravitational waves in flat space-time will become strongly coupled at energies lower than the Planck mass. This coupling is unnatural, and could even be considered unphysical. Notably, if the vector field is taken as a spectator, it also gives rise to the runaway behavior of modes, meaning that their gradient terms come with a wrong sign at high energies\footnote{One should note that this branch is different from the one in which the vector field has non-vanishing temporal component, and thus different cosmological structure as well as the overall number of degrees of freedom \cite{DeFelice:2025ykh}.} \cite{Capanelli:2024pzd}. 

However, as shown in \cite{Hell:2024xbv}, if one introduces disformal couplings to the theory, and couples it minimally to matter in this disformal frame, both runaway and the strong coupling of the tensor modes disappear. Such couplings in the form of metric transformations with inclusion of a term non-linear in the fields have been widely studied in the construction of new theories of gravity \cite{Bekenstein:1992pj, Gumrukcuoglu:2019ebp, Deffayet:2013tca, Kimura:2016rzw, Domenech:2018vqj, DeFelice:2019hxb, Jirousek:2022rym, Takahashi:2021ttd, Deruelle:2014zza, Fujita:2015ymn, Domenech:2023ryc, Alinea:2022ygr, Naruko:2019jzj, Heisenberg:2016eld, Papadopoulos:2017xxx, Bettoni:2013diz, Zumalacarregui:2013pma, Domenech:2015hka, Domenech:2015tca, Crisostomi:2016czh, BenAchour:2016cay, Domenech:2025eii, Gorji:2025ajb}.

If one introduces them to the theory of a  vector field with non-minimal coupling to gravity, the inconsistent strong coupling of the tensor modes, is replaced by the strong coupling of the scalar modes that are absent in the massless theory, making thus this effect a natural one. In addition, in the disformal frame, the runaway modes that are also absent. Given these surprising properties of the non-minimally coupled vector fields, and their resolution in the disformal frame, it is natural to pose a question if such effects can also be present for other fields, and resolved in a similar way? In this work, we will focus on exploring this question for the massive Kalb-Ramond field with non-minimal coupling to gravity. 

Kalb-Ramond (KR) field is an antisymmetric rank-2 tensor that naturally arises in string theory as a fundamental sector \cite{Ogievetsky:1966eiu, Kalb:1974yc}. 
It also naturally arises in theories of gravity such as in non-symmetric or hermitian gravity \cite{Einstein:1945eu, Moffat:1978tr, Chamseddine:2005at, Chamseddine:2006epa, Chamseddine:2012gh}, as well as gravity with antisymmetric components formulated in terms of the vierbein field \cite{Markou:2018wxo, Markou:2018mhq}.
It is very relevant for the cosmological and field theory studies, when it arises from theories of gravity \cite{Prokopec:2005fb, Prokopec:2006kr, Mantz:2008hm}, and also by being often regarded as a dual of an axion field \cite{Svrcek:2006yi}.
Owing to their couplings to other fields, KR fields can lead to a variety of cosmological phenomena.
For example, a dilatonic coupling in their kinetic term may drive particle production of the KR field, depending on the background dynamics of the dilaton during inflation.
Such dynamics have been studied in the context of anisotropic inflation, where they can induce an attractor solution \cite{Koivisto:2009sd, Ohashi:2013mka,Ohashi:2013qba,Ito:2015sxj}, as well as generate statistically anisotropic gravitational waves or primordial black holes after inflation \cite{Obata:2018ilf,Fujita:2022ait}.
Its non-minimal coupling to gravity has also been extensively studied, in the context of spontaneous violation of Lorentz symmetry \cite{Altschul:2009ae}, generation of dark matter through gravitational particle production \cite{Capanelli:2023uwv}, and parity-violating gravitational waves \cite{Manton:2024hyc, Alexander:2025wnj, Horii:2025jen}.

Given the recently unnatural appearance of runaway modes and strongly coupled tensor modes in the presence of vector fields with non-minimal coupling to gravity, it is worth to explore if the same will take place for the non-minimally coupled KR fields. In this work, we will show that this indeed takes place, as long as the KR field is coupled to the Ricci and Riemann tensors.

This paper is organized as follows: First in the Section \ref{section 2} we will discuss the basic building blocks of massive, minimally coupled KR field. We will then first explore their behavior in flat space-time in Section \ref{section 3}. This will be followed by the studies of the propagation of modes in homogeneous and isotropic background in Section \ref{section 4}, and the discussion on disformal couplings in the Section \ref{section 5}. Finally, we will discuss our results in the Section \ref{section 6}.

\section{The basics of Kalb-Ramond fields} \label{section 2}

The theory quadratic in Kalb-Ramond fields with non-minimal coupling to gravity and no higher derivatives is described by the following action:
\begin{equation}\label{full_action}
    \begin{split}
        S=\int d^4x\sqrt{-g}&\left\{\frac{\Mpl^2}{2}R-\frac{1}{12}H_{\mu\nu\rho}H^{\mu\nu\rho}-\frac{m^2}{4}B_{\mu\nu}B^{\mu\nu}-\alpha R B_{\mu\nu}B^{\mu\nu}-\zeta RB_{\mu\nu}\Tilde{B}^{\mu\nu}\right.\\&\left.
        -\left[\beta R^{\mu\nu}B_{\mu\alpha}B_{\nu}^{\;\;\alpha}+\gamma R^{\mu\nu\rho\sigma}B_{\mu\nu}B_{\rho\sigma}+\lambda\Tilde{R}^{\mu\nu\rho\sigma}B_{\mu\nu}B_{\rho\sigma}+\sigma R^{\mu\nu}B_{\mu\lambda}\Tilde{B}_{\nu}^{\;\;\lambda}\right]\right\}
    \end{split}
\end{equation}
Here, the tildes denote the dual fields: 
\begin{equation}
    \tilde{R}^{\mu\nu\rho\sigma}=\frac{1}{2}\varepsilon^{\mu\nu\alpha\beta}R_{\alpha\beta}^{\;\;\;\;\rho\sigma}\qquad\text{and}\qquad \tilde{B}^{\mu\nu}=\varepsilon^{\mu\nu\alpha\beta}B_{\alpha\beta},
\end{equation}
$B_{\mu\nu}=-B_{\nu\mu}$ is the anti-symmetric two form, known as the KR field, and $H_{\mu\nu\rho}$ is its corresponding field-strength tensor, given by:
\begin{equation}
    H_{\mu\nu\rho}=\nabla_{\mu}B_{\nu\rho}+\nabla_{\nu}B_{\rho\mu}+\nabla_{\rho}B_{\mu\nu},
\end{equation}
and $\varepsilon^{\mu\nu\alpha\beta}$ is the Levi-Civita tensor. 
Our goal in this work is to analyze the above theory from two perspectives: first, by studying the non-linear corrections of the above theory in the flat space-time, and second by examining the theory in the homogeneous and isotropic Universe, treating the KR field as a spectator field. In this section, we will set up the basis for these two goals, analyzing the KR theory in flat space-time. 

For this, let us consider the action of a free KR field: 
\begin{equation}\label{freeKR}
    S_{KR}=\int d^4x \left(-\frac{1}{12}H_{\mu\nu\rho}H^{\mu\nu\rho}-\frac{m^2}{4}B_{\mu\nu}B^{\mu\nu}\right),
\end{equation}
and analyze its degrees of freedom in the Minkowski space-time. Following the analysis similar to \cite{Chamseddine:2012gh, Hell:2021wzm}, we will decompose the field into irreducible representations of the rotation group:
\begin{equation}\label{decompositionKR}
    \begin{split}
        B_{0i}&\equiv C_i=C_i^T+\mu_{,i},\\
        B_{ij}&\equiv B_{ij}^T+\varepsilon_{ijk}\chi_{,k}=V_{j,i}^T-V_{i,j}^T+\varepsilon_{ijk}\chi_{,k},
    \end{split}
\end{equation}
where
\begin{equation}
    C_{i,i}^T=0\qquad\text{and}\qquad V_{i,i}^T=0,
\end{equation}
and $_{,i}=\partial_i$ denotes a derivative with respect to the $x^i$ coordinate. By substituting the above decomposition into the action (\ref{freeKR}), we find:
\begin{equation}
   \begin{split}
        \mathcal{L}_{KR}&=\frac{1}{2}\left[C_i^T(-\Delta+m^2)C_i^T+2C_i^T\Delta\dot{V}_i^T-\dot{V}_i^T\Delta \dot{V}_i^T+m^2V_i^T\Delta V_i^T\right]\\&
        -\frac{1}{2}\left[m^2\mu\Delta\mu+\dot{\chi}\Delta\dot{\chi}+\Delta\chi\Delta\chi-m^2\chi\Delta\chi\right],
   \end{split}
\end{equation}
where dot denotes the derivative with respect to the coordinate time, and $\Delta=\partial_i\partial_i$ is the Laplacian. We can see that in the above relation, the $C_i^T$ and $\mu$ are not propagating. The scalar satisfies the following constraint:
\begin{equation}
    \Delta\mu=0,
\end{equation}
whose solution is:
\begin{equation}
    \mu=0,
\end{equation}
while the vector satisfies:
\begin{equation}
    (-\Delta+m^2)C_i^T=-\Delta \dot{V}_i^T\qquad\to\qquad C_i^T=\frac{-\Delta}{-\Delta+m^2} \dot{V}_i^T. 
\end{equation}
By substituting the solution of the above constraint back to the action, we find the following Lagrangian density: 
\begin{equation}
    \mathcal{L}_{KR}=-\frac{1}{2}V_i^T(-\Box+m^2)\frac{-\Delta m^2}{-\Delta+m^2} V_i^T-\frac{1}{2}\chi(-\Box+m^2)(-\Delta\chi),
\end{equation}
where $\Box=\partial_{\mu}\partial^{\mu}$. We can notice right away two important points: there are three degrees of freedom -- two vector modes, and a pseudo-scalar, longitudinal one. In addition, the two vector modes are multiplied by a mass. Thus, when one sets $m=0$, they drop out from the action. This is also natural, because if massless, the KR field describes one degree of freedom. In addition, this also affects their minimal amplitude of quantum fluctuations. While for the canonically normalized scalar field in position space we find $\delta\phi_k\sim\sqrt{\frac{k^3}{\omega_k}}\sim\left.\frac{1}{L}\right|_{k\sim\frac{1}{L}}$, with $\omega_k = \sqrt{k^2 + m^2}$, for small length-scales, $\frac{1}{L}\gg m$ \cite{Mukhanov:2007zz}, this implies that:
\begin{equation}\label{qfluctKR}
    \delta V_i^T\sim\frac{1}{mL}\qquad \text{and}\qquad \delta\chi\sim\mathcal{O}(1).
\end{equation}
for the original modes. This will be important for the following section, when we compare the effects of the interactions and determine the strong coupling scale -- the scale at which the perturbation theory breaks down.

\section{The non-minimal coupling \textit{vs.} the strong coupling} \label{section 3}

In the previous section, we have seen the key properties of the Kalb-Ramond field, studying the free fields in the flat space-time. In this case, we have confirmed that the KR field describes three degrees of freedom -- the two transverse modes, and a longitudinal one, which in contrast to the former survives when one sets $m=0$. The goal of this section is to use these results, and study the KR theory in the presence of non-minimal coupling, while keeping the background flat. 

\subsection{The standard picture} 

One of the important points that can notice when considering the minimal amplitude of quantum fluctuation for the modes of the KR field in (\ref{qfluctKR}) is that the amplitude for the vector modes is singular in mass. To see what this means, let us consider the case of quartic self-interaction, which was discussed in detail in \cite{Hell:2021wzm}:
\begin{equation}
    \mathcal{L}_{KR}=-\frac{1}{12}H_{\mu\nu\rho}H^{\mu\nu\rho}-\frac{m^2}{4}B_{\mu\nu}B^{\mu\nu}-\frac{\lambda_{KR}}{16}\left(B_{\mu\nu}B^{\mu\nu}\right)^2,
\end{equation}
where $\lambda_{KR}\ll 1$ is the coupling constant. If one finds the constraints, solves them, and substitutes them back into the action, one will still remain with three degrees of freedom, which will now self-interact. For length-scales $\frac{1}{L^2}\gg m^2$, the equations of motion can be schematically represented as: 
\begin{equation}
    \begin{split}
        m^2(-\Box+m^2)V^T\sim \frac{\lambda_{KR}}{L^4} V^{T3} \qquad \text{and}\qquad (-\Box+m^2)\chi\sim \frac{\lambda_{KR}}{L^2}V^{T3}
    \end{split}
\end{equation}
to the leading order, 
 where we have replaced all of the derivatives with $\partial_{\mu}\sim\frac{1}{L}$ and suppressed all of the indices. Next, to solve the above equations, one can apply perturbation theory: 
\begin{equation}
    V_i^T=V_i^{T(0)}+V_i^{T(1)}+...\qquad \text{and}\qquad  \chi=\chi^{(0)}+\chi^{(1)}+..., 
\end{equation}
where the components with $(0)$ index satisfy the free equations. Thus, by taking into account (\ref{qfluctKR}), the first-order corrections can be estimated as:
\begin{equation}
    V_i^{(1)}\sim\frac{\lambda_{KR}}{ (mL)^5} \qquad \text{and}\qquad \chi^{(1)}\sim \frac{\lambda_{KR}}{(mL)^3}
\end{equation}
The first-order correction to the vector modes becomes of the same order as the linear term $V_i^{(0)}\sim\frac{1}{mL}$ at length scale:
\begin{equation}
    L_{V,str}\sim\frac{\lambda_{KR}^{1/4}}{m}, 
\end{equation}
whereas the correction for the longitudinal mode we find that $\chi^{(0)}\sim\chi^{(1)}$ at:
\begin{equation}
    L_{\chi,str}\sim\frac{\lambda_{KR}^{1/3}}{m}. 
\end{equation}
However, since, $L_{V,str}>L_{\chi,str}$, the perturbation theory breaks down as soon as one reaches the $L_{V,str}$ scale, making thus the scale $L_{\chi,str}$ is no longer trustable. Therefore, at this scale, the vector modes become strongly coupled. Notably, one can show that for this theory, they remain strongly coupled for $L<L_{V,str}$ and decouple from the longitudinal modes up to small corrections, which, in turn, remain in the weakly-coupled regime below and beyond the strong coupling scale \cite{Hell:2021wzm}. 

The previous behaviour is natural from the point of the principle of continuity \cite{BassSch} -- if one modifies the theory by a small parameter, one would naturally expect that all observables smoothly approach the original ones in the limit when this parameter tends to zero. By studying various massive gauge theories or modified theories of gravity, one can notice that a similar trend occurs: if a new dof appears upon modifying a theory, usually this mode becomes strongly coupled, and decouples from the remaining degrees of freedom in the limit when this parameter tends to zero. This was shown to be the case in massive gravity, where is known as the Vainshtein mechanism, and also applied to several other theories, including self-interacting Abelian vector fields, the massive Yang-Mills theory,  massive two-form and 3-form theories, massive mimetic gravity, as well as in $R^2$ gravity in flat space-time \cite{Vainshtein:1972sx, Deffayet:2001uk, Gruzinov:2001hp, Dvali:2007kt, DeFelice:2016cri, Chamseddine:2018gqh, Hell:2021wzm, Hell:2021oea,  Hell:2022wci, Hell:2023mph, Hell:2025uoc, Hell:2025pso}. 

Let us also compare this with Einstein's gravity, described by the action
\begin{equation}
    S=\frac{\Mpl^2}{2}\int d^4x\sqrt{-g}R,
\end{equation}
linearized around flat space-time: 
\begin{equation}
    g_{\mu\nu}=\eta_{\mu\nu}+h_{\mu\nu}. 
\end{equation}
In this case, the Lagrangian density is given by:
\begin{equation}
    \mathcal{L}=\frac{\Mpl^2}{8}\left(2h_{,\mu}^{\mu\nu}h^{\alpha}_{\nu,\alpha}-h_{\alpha\beta,\mu}h^{\alpha\beta,\mu}+2hh^{\mu\nu}_{,\mu\nu}+h_{,\mu}h^{,\mu}\right)
\end{equation}
By decomposing the metric according to the irreducible representations of the rotation group:
\begin{equation}\label{decomposition}
    \begin{split}
        &h_{00}=2\phi\\
        &h_{0i}=B_{,i}+S_i,\qquad\qquad S_{i,i}=0\\
        &h_{ij}=2\psi\delta_{ij}+2E_{,ij}+F_{i,j}+F_{j,i}+h_{ij}^{T},\qquad\qquad F_{i,i}=0,\quad h_{ij,i}^{T}=0,\quad h_{ii}^{T}=0
    \end{split}
\end{equation}
we find that at the leading order the scalars, vectors, and tensor decouple: 
\begin{equation}
    \mathcal{L}=\Mpl^2\left[2\phi\Delta\psi-3\dot{\psi}^2-\psi\Delta\psi+\frac{1}{4}S_{i,j}S_{i,j}-\frac{1}{8}h_{ij,\mu}^Th^{T,\mu}_{ij}\right]+\mathcal{O}(h^3)
\end{equation}
where we have chosen the longitudinal and Poisson gauges:
\begin{equation}
    E=0,\qquad B=0\qquad F_i=0. 
\end{equation}
We can notice that due to the $\phi$ and $S_i$ constraints:
\begin{equation}
    \psi=0\qquad \text{and}\qquad S_i=0. 
\end{equation}
Therefore, only the tensor modes propagate: 
\begin{equation}
    \mathcal{L}=\Mpl^2\left[-\frac{1}{8}h_{ij,\mu}^Th^{T,\mu}_{ij}\right]+\mathcal{O}(h^3)
\end{equation}
with the minimal amplitude of quantum fluctuations given by:
\begin{equation}\label{qfluctTenM}
    \delta h_{ij}^T\sim\frac{1}{\Mpl L}.
\end{equation}
Clearly, this is singular in $\Mpl$ and thus also agrees with the previous behaviour -- the Planck mass will dictate the scale at which the perturbation theory breaks down. To see this, we should take into account the second-order corrections, which arise schematically in the form:
\begin{equation}
    \Box h_{ij}^{T(1)}\sim \frac{(h^{T(0)})^2}{L^2}\sim\frac{1}{\Mpl^2L^4},
\end{equation}
where $(1)$ and $(0)$ refer to the first order correction in the perturbative theory, and the free part of the equation:
\begin{equation}
    \Box h_{ij}^{T(0)}=0. 
\end{equation}
These become of the same order as the linear term once one reaches the Planck mass:
\begin{equation}
    L\sim\frac{1}{\Mpl}. 
\end{equation}
Therefore, the characteristic strong-coupling scale for the tensor modes in the Minkowski space is the Planck mass.

\subsection{The non-minimal coupling}
Let us now generalize the previous analysis to the case of the non-minimal couplings to gravity. By expanding the action (\ref{full_action}) in the metric perturbations, we will encounter an infinite series of terms. However, not all of them are equally relevant. In particular, from the Einstein-Hilbert term, we will find metric perturbations at second, cubic, quartic and higher orders:
\begin{equation}
    \mathcal{O}\left(\frac{\Mpl^2}{L^2} h^3\right),\qquad   \mathcal{O}\left(\frac{\Mpl^2}{L^2} h^4\right), \qquad ...
\end{equation}
which contain the scalar, vector and tensor perturbations, now coupled in comparison to the linearized case. Among them, from the previous section, we know that the tensor modes are the propagating ones, while the scalar and vector modes are constrained. By taking into account (\ref{qfluctTenM}), the cubic and quartic higher-order terms involving the tensor modes can be estimated as:
\begin{equation}
    \mathcal{O}\left(\frac{\Mpl^2}{L^2} (h^T)^3\right)\sim\frac{1}{M_p L^5},\qquad   \mathcal{O}\left(\frac{\Mpl^2}{L^2} (h^T)^4\right)\sim\frac{1}{M_p^2 L^6}
\end{equation}
Therefore, if $L^{-1}<\Mpl$, the higher-order terms will be subdominant compared to the cubic term -- the first order correction. 

The higher-order corrections also arise from the kinetic term for the KR field. The leading order corrections in this case involve the interactions between the tensor modes and the KR field components: 
\begin{equation}\label{KRkintermsestimate}
    \mathcal{L}_{KR}=-\frac{1}{12}\sqrt{-g}\left(H_{\mu\nu\rho}H^{\mu\nu\rho}+3m^2B_{\mu\nu}B^{\mu\nu}\right)\qquad\to\qquad\frac{m^2}{L^2}h^T(V^T)^2,\quad \frac{1}{L^4}h^T\chi^2,\quad \frac{m^2}{L^2}h^T\chi^2, 
\end{equation}

In addition to the tensor modes, the KR field will also be coupled to the scalar and vector ones. However, once we find their constraints, solve them, and substitute back, we can verify that they will give rise to quartic self-interactions for the KR field, and not contribute at the leading order (cubic, on the level of action). The similar situation also takes place if one considers the coupling with the Ricci scalar. As we will see, at the leading order for $mL\ll 1$, it will yield self-interacting vector modes: 
\begin{equation}
    RB_{\mu\nu}B^{\mu\nu} \qquad \to \qquad V^{T4} 
\end{equation}
because the Ricci scalar at its leading order perturbation around the Minkowski space-time involves only the scalar modes. However, the same will not appear for the Ricci tensor, and the Riemann tensor, which involve tensor modes at leading order, which will couple to the KR field. As we will see, these interactions are dominant when compared to the higher-order terms arising from only the Einstein-Hilbert action, or the standard kinetic term of the KR field in (\ref{KRkintermsestimate}). Let us now study all of the non-minimal couplings to the leading order in the perturbations. To get the intuition about each of them, we will study them separately, taking into account only the most important terms in the expansion. 

In particular, we will now consider the Ricci scalar,  tensor and Riemann tensor couplings with the KR field, as well as their dual counterparts. To keep the overview of the interacitons clear, we will set all couplings to zero, apart from those that we are considering, and state only the dominant terms in the Lagrangian density. 

\subsubsection{The Ricci scalar coupling: type $\alpha$}

At the leading order in perturbations, the Lagrangian density that characterized the coupling between the Ricci scalar and the KR field is given by:
\begin{equation}
    \begin{split}
      \mathcal{L}=-\sqrt{-g}\alpha R B_{\mu\nu}B^{\mu\nu}=-\alpha (\partial_{\mu}\partial_{\nu}h^{\mu\nu}-\Box h)B_{\mu\nu}B^{\mu\nu}+\mathcal{O}\left(\alpha h^2B^2\right)
    \end{split}
\end{equation}

Therefore, the most relevant terms in the Lagrangian density, with all other non-minimal couplings set to zero are: 
\begin{equation}
    \begin{split}
        \mathcal{L}&=\mathcal{L}_0+\mathcal{L}_{int}\\
        \mathcal{L}_0&=\frac{1}{2}\left[C_i^T(-\Delta+m^2)C_i^T+2C_i^T\Delta\dot{V}_i^T-\dot{V}_i^T\Delta \dot{V}_i^T+m^2V_i^T\Delta V_i^T\right]\\&
        -\frac{1}{2}\left[m^2\mu\Delta\mu+\dot{\chi}\Delta\dot{\chi}+\Delta\chi\Delta\chi-m^2\chi\Delta\chi\right]\\
       &+ \Mpl^2\left[2\phi\Delta\psi-3\dot{\psi}^2-\psi\Delta\psi+\frac{1}{4}S_{i,j}S_{i,j}-\frac{1}{8}h_{ij,\mu}^Th^{T,\mu}_{ij}\right]\\
        \mathcal{L}_{int}&=-2\alpha\left(\Delta\phi+3\Ddot{\psi}-2\Delta\psi\right)B_{\mu\nu}B^{\mu\nu}+\mathcal{O}\left(\frac{m^2}{L^2}h^T(V^T)^2, \frac{1}{L^4}h^T\chi^2, \frac{\Mpl^2}{L^2} (h^T)^3\right),
    \end{split}
\end{equation}
where we have kept the KR field in the last line in the non-expanded form for simplicity. As the next step, we find the constraint for $\phi$, which is to the leading order given by:
\begin{equation}
    \Delta\psi\sim\frac{\alpha}{\Mpl^2}\Delta\left(B_{\mu\nu}B^{\mu\nu}\right). 
\end{equation}
We solve this for $\psi$, and substitute back to the action. Next, we find and solve the constraints for $S_i$, $C_i^T$ and $\mu$. With this, we obtain the Lagrangian density that contains only propagating modes, and is given to the leading order by: 
\begin{equation}
    \begin{split}
        \mathcal{L}_{\alpha}&=-\frac{\Mpl^2}{8}h_{ij,\mu}^Th_{ij}^{T,\mu}-\frac{1}{2}V_i^T(-\Box+m^2)\frac{(-\Delta)m^2}{-\Delta+m^2}V_i^T-\frac{1}{2}\chi(-\Box+m^2)(-\Delta\chi)\\
        &+\frac{12\alpha^2}{\Mpl^2}\left[V_{i,\mu}^TV^{T,\mu}_i-V_{i,j}^TV_{j,i}^T\right]\Box\left[V_{k,\nu}^TV^{T,\nu}_k-V_{k,l}^TV_{l,k}^T\right]\\&+\mathcal{O}\left(\frac{m^2}{L^2}h^T(V^T)^2, \frac{1}{L^4}h^T\chi^2, \frac{\Mpl^2}{L^2} (h^T)^3, \frac{\alpha^2}{\Mpl^2L^6}(V^T)^3\chi\right),
    \end{split}
\end{equation}
One should note that the above Lagrangian has many more terms. However, in the above, we have isolated only the most important ones, which are leading in the perturbative expansion. The equation of motion for the vector modes containing the most dominant contributions is given by:
\begin{equation}
   \begin{split}
        (-\Box+m^2)\frac{m^2(-\Delta)}{-\Delta+m^2}V_k^T&=-\frac{48\alpha^2}{\Mpl^2}P_{kl}^T\left[\partial^{\nu}\left(V_{l,\nu}\Box\left(V_{i,\mu}^TV^{T,\mu}_i-V_{i,j}^TV_{j,i}^T\right)\right)\right.\\&\left.-\partial_n\left(V_{n,l}\Box\left(V_{i,\mu}^TV^{T,\mu}_i-V_{i,j}^TV_{j,i}^T\right)\right)\right],
   \end{split}
\end{equation}
where
\begin{equation}
    P_{ij}^T=\delta_{ij}-\frac{\partial_i\partial_j}{\Delta}
\end{equation}
is the transverse projector. By expanding 
\begin{equation}
    V_i^T=V_i^{T(0)}+V_i^{T(1)}+..., 
\end{equation}
where
\begin{equation}
    (-\Box+m^2)V_i^{T(0)}=0,
\end{equation}
we find that the leading-order corrections can be estimated for $\frac{1}{L^2}\gg m^2$ as:
\begin{equation}
    V_i^{T(1)}\sim\frac{\alpha^2}{\Mpl^2m^5L^7}
\end{equation}
Therefore, one the length-scale:
\begin{equation}
    L_{\alpha,str}\sim\left(\frac{\alpha}{\Mpl m^2}\right)^{1/3}
\end{equation}
is reached, the vector modes become strongly coupled. This is natural -- they are also absent in the case of the massless theory. Thus, one could expect that  they also decouple from the remaining modes for the energies below this scale. 

\subsubsection{The Ricci scalar coupling: type $\zeta$ }
As the next step, let us consider the coupling with the Ricci scalar, that involves the dual of the KR field:
\begin{equation}
   \mathcal{L}=-\sqrt{-g}\zeta R B_{\mu\nu}\tilde{B}^{\mu\nu}
\end{equation}
In this case, the procedure is similar to the previous section. In particular, we first find that at the leading order, the constraint for the gravitational potential $\phi$ leads to 
\begin{equation}
    \Delta\psi\sim\frac{\zeta}{\Mpl^2}\Delta\left(B_{\mu\nu}\tilde{B}^{\mu\nu}\right). 
\end{equation}
By substituting this back into the Lagrangian density, and further resolving the constraints for $\mu$, $\xi_i$ and $C_i^T$, we find:
\begin{equation}
    \begin{split}
        \mathcal{L}_{\zeta}&\sim-\frac{\Mpl^2}{8}h_{ij,\mu}^Th_{ij}^{T,\mu}-\frac{1}{2}V_i^T(-\Box+m^2)\frac{(-\Delta)m^2}{-\Delta+m^2}V_i^T-\frac{1}{2}\chi(-\Box+m^2)(-\Delta\chi)\\&+\frac{192\zeta^2}{\Mpl^2}\varepsilon_{ijk}\varepsilon_{lmn}\dot{V}_i^TV_{k,j}^T\Box\left(\dot{V}_l^TV_{n,m}^T\right)      
    \end{split}
\end{equation}
at the leading order in the perturbation theory. Therefore, similarly to the previous case, we can find that the scale at which the vector modes become strongly coupled is given by;:
\begin{equation}
    L_{\xi,str}\sim\left(\frac{\zeta}{\Mpl m^2}\right)^{1/3}
\end{equation}

\subsubsection{The Ricci tensor coupling: type $\beta$}\label{betaC}

In the previous section we have seen that coupling the KR field to the Ricci scalar gives rise to the vector modes that are strongly coupled. This is also expected, when the theory is perturbatively analysed, the amplitude of quantum fluctuations for the vector modes is singular in mass, whereas that for the pseudo-scalar is well-behaved as far as the mass is concerned. In contrast, the tensor modes should be strongly coupled once the energies reach the Planck mass -- indicated by their corresponding amplitude of quantum fluctuations. 

Let us now see how the things change if one considers the coupling with the Ricci tensor. In this case the Lagrangian density describing the non-minimal coupling is given by:
\begin{equation}
    \mathcal{L}=-\sqrt{-g}\beta R^{\mu\nu}B_{\mu\alpha}B_{\nu}^{\;\;\alpha}
\end{equation}
To the leading order in the perturbations, the above term is given by:
\begin{equation}
    \mathcal{L}=-\beta \left[h^{\mu\beta,\nu}_{,\beta}-\frac{1}{2}\left(\Box h^{\mu\nu}+h^{,\mu,\nu}\right)\right]B_{\mu\alpha}B_{\nu}^{\;\;\alpha}
\end{equation}
By decomposing the fields, one then finds: 
{\begin{equation}
    \begin{split}
     \mathcal{L}_{\beta int}=&-\frac{\beta}{2}\left\{\phi\left[-2\Delta(C_iC_i)-2\partial_i\partial_j(C_iC_j)+2\partial_i\partial_j(B_{il}B_{jl})\right]\right.\\&\left. 
     +\psi\left[-8\partial_0^2(C_iC_i)+2\Delta(C_iC_i)+2\partial_i\partial_j(C_iC_j)+8\partial_i\partial_0(C_jB_{ij})-2\partial_i\partial_j(B_{il}B_{jl})-2\Box(B_{il}B_{il})\right]\right.\\&\left. 
     +S_j\left[2\partial_i\partial_0(C_iC_j)+2\Delta(C_iB_{ji})-2\partial_i\partial_0(B_{il}B_{jl})\right]\right.\\&\left. 
     +\Box h_{ij}^T\left[C_iC_j-B_{il}B_{jl}\right]\right\}
    \end{split}
\end{equation}
}

In the following, we will be interested in the leading terms for this theory. To find them, we repeat the procedure as in the previous cases -- we find the constraint for $\phi$, which can be solved for the scalar $\psi$. Then, by substituting this back to the Lagrangian, we find that also $\mu$, $S_i$ and $C_i$ do not propagate as well, while $\phi$ drops out of the action since the constraint in it is linear. After finding and resolving the constraints for the remaining fields, $\mu$, $S_i$ and $C_i$, solving them, and substituting back to the action, we find the Lagrangian density, expressed only in terms of the propagating modes. 
Its form, including only the most dominant terms {in the limit $1/L^2 \gg m^2$} is given by

\begin{equation}
    \begin{split}
        \mathcal{L}_{\beta}&\sim-\frac{\Mpl^2}{8}h_{ij,\mu}^Th_{ij}^{T,\mu}-\frac{1}{2}V_i^T(-\Box+m^2)\frac{(-\Delta)m^2}{-\Delta+m^2}V_i^T-\frac{1}{2}\chi(-\Box+m^2)(-\Delta\chi)\\&-\frac{\beta}{2}\left[\dot{V}_i^T\dot{V}_j^T-(V_{l,i}^T-V_{i,l}^T)(V_{l,j}^T-V_{j,l}^T)\right]\Box h_{ij}^T+\frac{\beta^2}{\Mpl^2}\frac{(V^T)^4}{L^6}\\
        &+\mathcal{O}\left(\frac{\Mpl^2}{L^2}(h^T)^3, \frac{m^2}{L^2}h^T(V^T)^2,\frac{h^T\chi^2}{L^4}\right)
    \end{split}
\end{equation}
The last term in the second row is the schematic expression for all the quartic vector self-interactions that appear at the second order in the perturbations. Similarly to the case with the Ricci scalar, this term will affect the leading order corrections to the vector modes, and therefore give rise to the strong coupling scale:
\begin{equation}
    L_{\beta,str}\sim\left(\frac{\beta}{\Mpl m^2}\right)^{1/3}. 
\end{equation}
However, in contrast to the previous case, this theory also involves the coupling between the tensor and vector modes, appearing at the cubic order. As a result, the equation of motion for the tensor modes becomes:
\begin{equation}
    \Box h^{T}_{kl}\sim\frac{2\beta}{\Mpl^2}P^T_{kl,ij}\Box\left(\dot{V}_i^T\dot{V}_j^T-(V_{l,i}^T-V_{i,l}^T)(V_{l,j}^T-V_{j,l}^T)\right),
\end{equation}
where 
\begin{equation}
   \begin{split}
        P_{mn,ij}^T&=\frac{1}{2}\left(\delta_{im}\delta_{jn}+\delta_{jm}\delta_{in}\right)-\frac{1}{2}\delta_{ij}\left(\delta_{mn}-\frac{\partial_k\partial_l}{\Delta}\delta_{km}\delta_{ln}\right)\\
        &+\frac{1}{\Delta}\left[\frac{1}{2}\delta_{mn}\partial_i\partial_j+\frac{1}{2}\delta_{mk}\delta_{ln}\frac{\partial_i\partial_j\partial_k\partial_l}{\Delta}-\delta_{im}\delta_{ln}\partial_l\partial_j-\delta_{jm}\delta_{ln}\partial_l\partial_i\right]
   \end{split}
\end{equation}
is the transverse, traceless projector. 
The vector modes affect the first order corrections to the tensor modes such that they also become singular in mass:
\begin{equation}
    h_{ij}^{(1)T}\sim\frac{\beta}{L^4m^2\Mpl^2}
\end{equation}
They become of the same order as the linear term $h_{ij}^{(0)T}$ at exactly the same strong coupling scale as the longitudinal modes, $L_{\beta,str}$.

\subsubsection{The Riemann tensor coupling: type $\gamma$}
In addition to the coupling with the Ricci tensor, the 2-form field also allows for the coupling with the Riemann tensor: 
\begin{equation}
    \mathcal{L}=-\sqrt{-g}\gamma R^{\mu\nu\rho\sigma}B_{\mu\nu}B_{\rho\sigma}, 
\end{equation}
which gives rise to the following coupling at the leading order:
\begin{equation}
     \mathcal{L}= 2\gamma B_{\mu\nu}B_{\alpha\beta}h^{\alpha\mu,\nu,\beta},
\end{equation}
which becomes:
\begin{align}
\mathcal{L}_{\gamma} &= 2\gamma h^{\mu\rho,\nu\sigma}B_{\mu\nu}B_{\rho\sigma} \\
&= 2\gamma\left\{ 2\phi\partial_i\partial_j(C_iC_j) \right. \notag \\
&\left.-2S_j\left[\partial_0\partial_i(C_iC_j) + \partial_i\partial_k(C_iB_{jk})\right] \right. \notag \\
&\left.+2\psi\left[ \partial_0^2(C_iC_i) + 2\partial_0\partial_k(C_i B_{ik}) + \partial_k\partial_\ell(B_{ik}B_{il}) \right] \right. \notag \\
&\left.+h^T_{ij}\left[ \partial_0^2(C_iC_j) - 2\partial_0\partial_k(C_iB_{jk}) + \partial_k\partial_\ell(B_{ik}B_{j\ell})\right] \right\} \ .
\end{align}
After resolving all of the constraints following a procedure equivalent to the previous subsections, we find the following Lagrangian density for $k^2\sim\frac{1}{L^2}\gg m^2$ at the leading order 
\begin{equation}
    \begin{split}
        \mathcal{L}_{\gamma}&\sim-\frac{\Mpl^2}{8}h_{ij,\mu}^Th_{ij}^{T,\mu}-\frac{1}{2}V_i^T(-\Box+m^2)\frac{(-\Delta)m^2}{-\Delta+m^2}V_i^T-\frac{1}{2}\chi(-\Box+m^2)(-\Delta\chi)\\& +2\gamma\left[\dot{V}_i^T\dot{V}_{j}^T\Ddot{h}_{ij}^T-2\dot{V}_i^T(V^T_{k,j}-V^T_{j,k})\dot{h}_{ij,k}^T+h^T_{ij,k,l}(V^T_{i,k}-V^T_{k,i})(V^T_{j,l}-V^T_{l,j})\right]+\frac{\gamma^2}{\Mpl^2}\frac{(V^T)^4}{L^6}\\
        &+\mathcal{O}\left(\frac{\Mpl^2}{L^2}(h^T)^3, \frac{m^2}{L^2}h^T(V^T)^2,\frac{h^T\chi^2}{L^4}\right)
    \end{split}
\end{equation}
where the last term in the second row denotes the quartic order self-interactions in a schematic way. We can notice that just like in the case of the Ricci tensor, the tensor modes will be strongly coupled, in addition to the vector modes. Their equation of motion: 
\begin{equation}
    \Box h_{mn}^T=-\frac{8\gamma}{\Mpl^2}P^T_{mnij}\left(\partial_0^2(\dot{V}_i^T\dot{V}_{j}^T)-2\partial_0\partial_k(\dot{V}_i^T(V^T_{k,j}-V^T_{j,k}))+\partial_k\partial_l[(V^T_{i,k}-V^T_{k,i})(V^T_{j,l}-V^T_{l,j}]\right)
\end{equation}
gives rise to first-order corrections: 
\begin{equation}
    h_{ij}^{(1)T}\sim\frac{\gamma}{L^4m^2\Mpl^2}
\end{equation}
which are of the similar order as in the other cases, and singular in mass. Once they become of the same order as the linear the linear terms, the perturbation theory for the tensor modes is no longer valid. This correspods to the strong coupling scale: 
\begin{equation}
    L_{\gamma,str}\sim\left(\frac{\gamma}{\Mpl m^2}\right)^{1/3}. 
\end{equation}
at which also the vector modes become strongly coupled, due to the quartic order corrections. 

\subsubsection{The Ricci tensor coupling: type $\sigma$}

It is easy to see that the couplings with the Ricci tensor with the presence of Levi-Civita tensor will have a similar effect on the tensor modes, as in the case with the $\beta$ coupling. 

In particular, the interacting part of the Lagrangian density in this case is given by:
\begin{equation}
    \mathcal{L}=-\sigma R^{\mu\nu}B_{\mu\alpha}\tilde{B}_{\nu}^{\;\;\alpha}
\end{equation}

In this case, we follow the procedure similar to the $\beta$ coupling, and solve all of the constraints. This leads us to the following Lagrangian density, with the most important terms is given by:
\begin{equation}
    \begin{split}
        \mathcal{L}_{\sigma}&\sim-\frac{\Mpl^2}{8}h_{ij,\mu}^Th_{ij}^{T,\mu}-\frac{1}{2}V_i^T(-\Box+m^2)\frac{(-\Delta)m^2}{-\Delta+m^2}V_i^T-\frac{1}{2}\chi(-\Box+m^2)(-\Delta\chi)\\&-\varepsilon_{ilk}\sigma\dot{V}_k^T(V_{l,k}^T-V_{l,j}^T)\Box h_{ij}^T
    \end{split}
\end{equation}
Here, we can notice that there is again an interaction between the tensor modes and the vector modes at the cubic order. This leads us to the strong coupling scale, analogous to the $\beta$ coupling case: 

\begin{equation}
    L_{\sigma,str}\sim\left(\frac{\sigma}{\Mpl m^2}\right)^{1/3}. 
\end{equation}

\subsubsection{The Riemann tensor coupling: type $\lambda$}
Finally, let us consider the last possibility -- coupling of the type $\lambda$, for which the Lagrangian density takes the form:
\begin{equation}
    \mathcal{L}_{int}=-\lambda\tilde{R}^{\mu\nu\rho\sigma}B_{\mu\nu}B_{\rho\sigma}
\end{equation}
At the leading order in perturbations, this becomes: 
\begin{equation}
    \mathcal{L}_{int}=-2\varepsilon^{\rho\sigma\alpha\beta}B_{\mu\nu}B_{\rho\sigma}\partial_{\alpha}\partial^{\nu}h_{\beta}^{\mu},
\end{equation}
leading us to the following Lagrangian density after all constraints are resolved:
\begin{equation}
\begin{split}
        \mathcal{L}_{\lambda}&\sim-\frac{\Mpl^2}{8}h_{ij,\mu}^Th_{ij}^{T,\mu}-\frac{1}{2}V_i^T(-\Box+m^2)\frac{(-\Delta)m^2}{-\Delta+m^2}V_i^T-\frac{1}{2}\chi(-\Box+m^2)(-\Delta\chi)\\&-\varepsilon_{nlu}\lambda\left[\dot{V}^T_i\dot{V}_l^T\dot{h}^T_{ui,n}+\dot{V}^T_l\dot{V}_j^T\dot{h}^T_{uj,n}+2\dot{V}_l^TV_{ij}^Th_{uj,sn}^T+2\dot{V}_j^TV_{nl}^T\ddot{h}_{uj}^T+2V_{js}^TV_{nl}^T\dot{h}_{uj,s}^T\right]
    \end{split}
\end{equation}
which leads us to the following strong coupling scale for the tensor modes:
\begin{equation}
    L_{\lambda,str}\sim\left(\frac{\lambda}{\Mpl m^2}\right)^{1/3}. 
\end{equation}

Therefore, overall, we find the following result: If one consider a KR field in flat space-time with non-minimal coupling to gravity only the coupling between a two-form and a Ricci scalar gives rise to the strong coupling of the KR field, which does not affect the remaining pseudo scalar or tensor modes. For the other non-minimal couplings, no matter how small or large the KR mass is, we find that the tensor modes will behave unnaturally, as they will become strongly coupled on the scales before the Planck mass is reached. 

\section{Mode propagation in the cosmological background} \label{section 4}

In the previous section we have considered the behavior of the Kalb-Ramond field in the presence of non-minimal couplings in flat space-time. In this case, we have found that such interactions contribute at the non-linear level, and give rise to unexpected properties in the tensor-mode sector, leading to their strong coupling, unless the only non-minimal coupling is that with the Ricci scalar. 

The goal of this section is to investigate another aspect of these couplings, when the flat space-time is replaced with the curved one. To present this in a pedagogical way, we will thus first present a free KR field in the homogeneous and isotropic background, and then generalize this study also to the presence of non-minimal couplings. 

\subsection{The minimal theory}
As a first step, let us assume that the background corresponds to the homogeneous and isotropic Universe, given by the Friedmann–Lema\^itre–Robertson–Walker (FLRW) metric: 
\begin{equation}
    ds^2=-dt^2 + a(t)^2d\bm{x}^2 = a(\tau)^2(-d\tau^2 + d\bm{x}^2),
\end{equation}
where $\tau$ stands for the conformal time. In this case, the action for the KR field becomes:
\begin{align}
S=&\frac{1}{2}\int d\tau d\bm{x} \dfrac{1}{a^2}\left[B_i{'}B_i{'} - 2\epsilon_{ijk}B_{0i}B'_{k,j} + B_{0i,j}B_{0i,j} - B_{0i,i}B_{0i,i} -B_{i,i}B_{i,i}\right.\notag \\
&\left. + a^2m^2(B_{0i}B_{0i}-B_kB_k)\right] \ ,
\end{align}
where the prime denotes the derivative with respect to the conformal time. By decomposing further the KR field according to the spatial rotations as in (\ref{decompositionKR}), we find:
\begin{align}
S = \dfrac{1}{2}\int d\tau d\bm{x} \dfrac{1}{a^2}&\left[ C^T_i\left(-\Delta + a^2m^2\right)C^T_i + 2C^T_i\Delta V{^{T}}'_{i} - a^2m^2\mu\Delta\mu \right. \notag \\
&\left. +  V{^{T}}'_{i,j}V{^{T}}'_{i,j} - a^2m^2V^T_{i,j}V^T_{i,j} - \phi'\Delta\phi' - \Delta\phi\Delta\phi + a^2m^2\phi\Delta\phi \right] \ .
\end{align}
We can notice that similarly to the case with flat space-time, two fields are not propagating: the vector mode $C_i^T$, and the scalar $\mu$. By varying the action with respect to them, we find the following two constraints:
\begin{align}
    \left(-\Delta + a^2m^2\right)C_i^T=-\Delta V{^{T}}'_i,\qquad\text{and}\qquad \Delta\mu=0,
\end{align}
whose solution is given by:
\begin{equation}
    C_i^T=\frac{-\Delta}{\left(-\Delta + a^2m^2\right)}V{^{T}}'_i\qquad\text{and}\qquad \mu=0
\end{equation}
By substituting these constraints back to the action, we can express it only in terms of the propagating modes:
    \begin{align}
S = \dfrac{1}{2}\int d\tau d\bm{x}&\left[ V{^{T}}'_i\dfrac{(-m^2\Delta)}{-\Delta + a^2m^2}V{^{T}}'_i + m^2V^T_i\Delta V^T_i  - \dfrac{1}{a^2}\phi'\Delta\phi' - \dfrac{1}{a^2}\Delta\phi\Delta\phi + m^2 \phi\Delta\phi \right] \ .
\end{align}
We can notice that similarly to the flat space-time, the theory describes two vector modes and a pseudo-scalar one. Let us now see how this generalizes if one considers non-minimal couplings. 

\subsection{The runaway modes }

In the presence of non-minimal coupling with spatially flat FLRW background the action (\ref{full_action}) becomes: 
\begin{align}
S=&\frac{1}{2}\int d\tau d\bm{x} \dfrac{1}{a^2}\left[B_i{'}^2 - 2\epsilon_{ijk}B_{0i}B'_{k,j} - 2a^2m_\lambda^2 B_{0i}B_{i}  + B_{0i,j}^2 - B_{0i,i}^2 -B_{i,i}^2 + a^2(m_t^2B_{0i}^2-m_s^2B_k^2)\right] \ ,
\end{align}
where we have defined
\begin{align}
m_t^2 &\equiv m^2 + \left(4\alpha + \beta + \dfrac{4}{3}\gamma\right)R - 2(3\beta + 4\gamma)H^2  \ , \\
m_s^2 &\equiv m^2 + \left(4\alpha + \dfrac{2}{3}\beta\right)R + 4(\beta + 2\gamma)H^2 \ , \\
m_\lambda^2 &\equiv \left(\dfrac{2}{3}\lambda + \sigma + 4\zeta\right) R
\end{align}
{with Hubble parameter $H \equiv \dot{a}/{a}$.}
One should note that in the above we have assumed that the background value of the KR field vanishes. This is also consistent with the homogeneous and isotropic background, as this property of the background would otherwise be violated, unless one considers multiple fields, such as a triplet, with an internal rotational symmetry.

By further decomposing the KR field as in (\ref{decompositionKR}), using the transverse conditions and partial integrations, the quadratic action for the 2-form perturbations is given by: 
\begin{align}
S = \dfrac{1}{2}\int d\tau d\bm{x} \dfrac{1}{a^2}&\left[ C^T_i\left(-\Delta + a^2m_t^2\right)C^T_i + 2C^T_i\Delta V{^{T}}'_{i} - 2a^2m_\lambda^2 (C^T_i\epsilon_{ijk}V^T_{k,j} -\mu\Delta\phi) - a^2m_t^2\mu\Delta\mu \right. \notag \\
&\left. +  V{^{T}}'_{i,j}V{^{T}}'_{i,j} - a^2m_s^2V^T_{i,j}V^T_{i,j} - \phi'\Delta\phi' - \Delta\phi\Delta\phi + a^2m_s^2\phi\Delta\phi \right] \ ,
\end{align}
where the dash symbol is a derivative with conformal time.
The non-dynamical variables are solved as
\begin{equation}
C^T_i = \dfrac{1}{-\Delta + a^2m_t^2}\left(-\Delta V{^T}'_{i} + a^2m_\lambda^2\epsilon_{ijk}V^T_{k,j}\right) \ , \qquad\text{and,}\qquad \Delta\mu = \dfrac{m_\lambda^2}{m_t^2}\Delta\phi \ .
\end{equation}
After integrating the non-dynamical variables out, we obtain
\begin{align}
S = \dfrac{1}{2}\int d\tau d\bm{x}&\left[ V{^{T}}'_i\dfrac{(-m_t^2\Delta)}{-\Delta + a^2m_t^2}V{^{T}}'_i + V^T_{k,j}\left(\dfrac{m_\lambda^2}{-\Delta+a^2m_t^2}\right)'\epsilon_{ijk}(-\Delta)V^T_i \right. \notag \\
&\left. - V^T_i\left(m_s^2 + \dfrac{a^2m_\lambda^4}{-\Delta + a^2m_t^2}\right)(-\Delta)V^T_i - \dfrac{1}{a^2}\phi'\Delta\phi' - \dfrac{1}{a^2}\Delta\phi\Delta\phi + \left(m_s^2 + \dfrac{m_\lambda^4}{m_t^2}\right)\phi\Delta\phi \right] \ .
\end{align}
Then, defining the following canonical variables
\begin{equation}
V^T_{ni} = \sqrt{\dfrac{(-m_t^2\Delta)}{-\Delta + a^2m_t^2}}V^T_i \equiv z_VV^T_i \ , \qquad \phi_n = \dfrac{1}{a}\sqrt{-\Delta}\phi \ ,
\end{equation}
and using the partial integration, the action reads
\begin{align}
S = \dfrac{1}{2}\int d\tau d\bm{x}&\left[ V{^{T}}'_{ni}V{^{T}}'_{ni} + \dfrac{z_\lambda'}{z_V^2}\epsilon_{ijk}V^T_{nk,j}(-\Delta)V^T_{ni} - \left(-\dfrac{m_s^2}{m_t^2}\Delta + a^2M_s^2  - \dfrac{z_V''}{z_V}\right)V^T_{ni}V^T_{ni} \right. \notag \\
&\left. + \phi_n'\phi_n' - \left(-\Delta + a^2M_s^2 - \dfrac{(1/a)''}{1/a}\right)\phi_n\phi_n \right] \ ,
\end{align}
where
\begin{equation}
M_s^2 \equiv m_s^2 + \dfrac{m_\lambda^4}{m_t^2} \ , \qquad \text{and,}\qquad z_\lambda \equiv \dfrac{m_\lambda^2}{-\Delta+a^2m_t^2} \ .
\end{equation}
Let us decompose it into Fourier modes.
Then, using the identity for the circular polarization vector $\epsilon_{ijk}ik_j e^\pm_{k}(\hat{\bm{k}}) = \pm k e^\pm_{i}(\hat{\bm{k}})$, we obtain
\begin{align}
S = \dfrac{1}{2}\int d\tau \dfrac{d\bm{k}}{(2\pi)^3}&\left[ |V{^{\pm}}'_{n,\bm{k}}|^2 - \left(\dfrac{m_s^2}{m_t^2}k^2 \mp \dfrac{z_\lambda'}{z_V^2}k^3 + a^2M_s^2 - \dfrac{z_V''}{z_V}\right)|V^\pm_{n,\bm{k}}|^2 \right. \notag \\
&\left. + |\phi_{n,\bm{k}}'|^2 - \left(k^2 + a^2M_s^2 - \dfrac{(1/a)''}{1/a}\right)|\phi_{n,\bm{k}}|^2 \right] \ .
\end{align}

Therefore, similarly to the minimal theory, we find three modes -- two vectors, and one pseudo-scalar. However, in contrast to it, the non-minimal coupling gives rise also to the effective mass. 

We can notice that the parity-violating coupling gives rise to a modified dispersion relation for the vector modes. However, this does not dominate at high-energies due to the $k^2$ factors appearing in $z_V$ making thus the overall scaling of this term not cubic but linear in the momenta.  In addition, the other couplings contribute too. To understand their contribution better, it is instructive to set the couplings $\lambda$, $\sigma$ and $\xi$ for the moment to zero. In this case, the action becomes:
\begin{align}
S = \dfrac{1}{2}\int d\tau \dfrac{d\bm{k}}{(2\pi)^3}&\left[ |V{^{\pm}}'_{n,\bm{k}}|^2 - \left(\dfrac{m_s^2}{m_t^2}k^2  + a^2m_s^2- \dfrac{z_V''}{z_V} \right)|V^\pm_{n,\bm{k}}|^2  + |\phi_{n,\bm{k}}'|^2 - \left(k^2 + a^2m_s^2 - \dfrac{(1/a)''}{1/a}\right)|\phi_{n,\bm{k}}|^2 \right] \ ,
\end{align}
with
\begin{align}
m_t^2 = m^2 + \left(4\alpha + \beta + \dfrac{4}{3}\gamma\right)R - 2(3\beta+4\gamma)H^2  \ , \\
m_s^2 = m^2 + \left(4\alpha + \dfrac{2}{3}\beta\right)R + 4(\beta + 2\gamma)H^2 \ .
\end{align}
We can notice that due to the presence of the Ricci and Riemann tensors, the high-k contribution for the vector modes becomes multiplied by the ratio $\dfrac{m_s^2}{m_t^2}$ which would otherwise be unity. We need to require that $m_t^2>0$, as otherwise the two vector modes would be ghosts. Therefore, for $m_t^2>0$ and $m_s^2<0$, it is possible give rise to a gradient instability known as the runaway. This effect for the KR field appears for the transverse vector modes. The longitudinal one, in contrast becomes tachyonic.  

It is curious to compare these results with a vector field with non-minimal coupling to gravity. In contrast to the KR field, a single vector field can have a non-vanishing background value in the presence of non-minimal coupling, and thus give rise to non-trivial cosmological evolution (see eg. \cite{DeFelice:2025ykh}). However, the theory can also admit a branch where the background value of the vector field vanishes. In this case, the theory gives rise to the runaway mode \cite{Capanelli:2024pzd} (see also \cite{Hell:2024xbv}). Similarly to the KR field, the vector field has three degrees of freedom -- two vector modes and a longitudinal one. Moreover, in \cite{Capanelli:2023uwv} it was claimed that the two theories with non-minimal coupling are dual, meaning that they describe the same physics. From this perspective, we can clearly see that this is not the case: not only the vector field admits background configurations that contribute to the cosmological expansion which is not possible to realize with the KR field, but in the branch of the theory where its background value vanishes, only its scalar mode admits the possibility for a gradient instability. In contrast, the vector modes in Proca theory can only be tachionic, whereas in the KR theory they can have a gradient instability. This difference in the behavior of the two theories clearly shows that the two theories do not describe the physics, further supporting the indication that the Proca and KR theories are not dual, as pointed originally pointed out in \cite{Hell:2021wzm} in the context of self-interactions. 

Interestingly, however, the parity-violating terms do not contribute to the runaway problem.
{However, there might exist a tachyonic instability for a certain momentum mode. To make it simple, we set $\alpha=\beta=\gamma=0$.
Then, $m_s^2=m_t^2 = m^2$ and we obtain
\begin{align}
S = \dfrac{1}{2}\int d\tau \dfrac{d\bm{k}}{(2\pi)^3}&\left[ |V{^{\pm}}'_{n,\bm{k}}|^2 - \left(k^2 \mp \dfrac{z_\lambda'}{z_V^2}k^3 + a^2M_s^2 - \dfrac{z_V''}{z_V}\right)|V^\pm_{n,\bm{k}}|^2 \right. \notag \\
&\left. + |\phi_{n,\bm{k}}'|^2 - \left(k^2 + a^2M_s^2 - \dfrac{(1/a)''}{1/a}\right)|\phi_{n,\bm{k}}|^2 \right] \ .
\end{align}
Focusing on the effective mass term for $V^\pm_{n,\bm{k}}$ {and $\phi_{n,\bm{k}}$}, it reads
\begin{align}
&k^2 \mp \dfrac{z_\lambda'}{z_V^2}k^3 + a^2M_s^2 - \dfrac{z_V''}{z_V} \notag \\
&= k^2\left[ 1 \pm \dfrac{2a^2m_\lambda^2}{k^2 + a^2m^2}\dfrac{a'}{ka} + \dfrac{a^2}{k^2}\left(m^2+\dfrac{m_\lambda^4}{m^2}\right) + \dfrac{a^2m^2}{k^2(k^2+a^2m^2)^2}\left[ (k^2-2a^2m^2)\dfrac{a'^2}{a^2}+(k^2+a^2m^2)\dfrac{a''}{a} \right] \right] \ , \label{eq: coe} \\
&k^2 + a^2M_s^2 - \dfrac{(1/a)''}{1/a} = k^2\left[ 1 + \dfrac{a^2}{k^2}\left(m^2+\dfrac{m_\lambda^4}{m^2}\right) - \dfrac{1}{k^2}\left( 2\dfrac{a'^2}{a^2} - \dfrac{a''}{a} \right) \right] \ . \label{eq: cop}
\end{align}

Let us consider the time evolution of KR field in an inflationary period. During inflation, $H \simeq \text{const.} \ (R \simeq 12H^2)$ and $a \simeq -(H\tau)^{-1}$.
Defining $2\lambda/3+\sigma+4\zeta \equiv y$, \eqref{eq: coe} and \eqref{eq: cop} read
\begin{align}
\eqref{eq: coe} &\simeq k^2\left[1 \pm \dfrac{24y}{k^2\tau^2 + m^2/H^2}\dfrac{1}{(-k\tau)} + \dfrac{1}{k^2\tau^2}\left( \dfrac{m^2}{H^2} + \dfrac{144H^2y^2}{m^2} \right) + \dfrac{3m^2}{H^2(k^2\tau^2+m^2/H^2)^2}\right] \ , \label{eq: coe2} \\
\eqref{eq: cop} &\simeq k^2\left[ 1 + \dfrac{1}{k^2\tau^2}\left( \dfrac{m^2}{H^2} + \dfrac{144H^2y^2}{m^2} \right) \right] \ . \label{eq: cop2}
\end{align}
We can see that tachyonic instability does not occur for $\phi_{n,\bm{k}}$.
For $V^\pm_{n,\bm{k}}$, it could occur when the second term becomes dominant and total value becomes negative.
To check the instability band, we solve the quadratic equation of \eqref{eq: coe2}=0 for $y$:
\begin{align}
y = \dfrac{-k\tau}{12}\dfrac{m^2}{H^2}\left[\dfrac{\mp1}{k^2\tau^2 + m^2/H^2} \mp 
   \sqrt{-\dfrac{1}{k^2\tau^2} - \dfrac{H^2}{m^2} - \dfrac{2}{(k^2\tau^2 + m^2/H^2)^2}}\right] \ .
\end{align}
One can find that there is no real solution: namely, tachyonic instability does not occur.
Thus, vector modes are diluted away during inflation.
One should note nevertheless, that the instability could take place during the radiation or matter dominated era after inflation, or if one assumes that $\alpha\neq0,\; \beta\neq0, \;\text{and}\; \gamma\neq0$.
Dedicated studies to explore the parameter space of such instability regimes and the associated cosmological phenomena are left for future work.

\section{Disformal couplings} \label{section 5}

So far, we have seen that the Kalb-Ramond theory with non-minimal coupling to gravity suffers from two issues. In flat space-time, its coupling with the Ricci and Riemann tensors including their duals gives rise to the strong coupling of the tensor modes, along with the vector modes. In addition, in the homogeneous and isotropic background, the two parity-preserving couplings give rise to the runaway modes in the vector sector, with gradient instability.  These two issues are entirely separate physical effects. However, it is natural to ask if they can be cured in the same way. 

Recently, in \cite{Hell:2024xbv} it was noticed that a similar effect occurs for the Proca field with non-minimal coupling to the Ricci tensor. It was found that in addition to the strong coupling of the longitudinal mode, no matter how small the mass of the vector field is, the two tensor modes would become strongly coupled as well, yielding a physically inconsistent result. Moreover, it was shown that the reason for the runaway modes in the scalar sector in the homogeneous and isotropic background is precisely due to the same coupling with the Ricci tensor. However, despite the two pathologies being of different physical origin, it was found that the inclusion of the disformal coupling can remove both effects. Curiously, the disformal coupling also has a non-Abelian analogue. Instead of metric introducing higher order terms, the non-Abelian field gets modified in the non-linear. Notably, inclusion of disformal couplings in this form is necessary if one aims to express the Yang-Mills theory in the gauge-invariant variables \cite{Hell:2022wci, Hell:2025pso}, and was also crucial to show that the massless limit of massive Yang-Mills theory where the mass is explicitly introduced by hand to the Lagrangian density instead of being generated through the Brout-Englert-Higgs mechanism is smooth \cite{Hell:2021oea}.  It has been widely used in the literature \cite{Bekenstein:1992pj, Gumrukcuoglu:2019ebp, Deffayet:2013tca, Kimura:2016rzw, Domenech:2018vqj, DeFelice:2019hxb, Jirousek:2022rym, Takahashi:2021ttd, Deruelle:2014zza, Fujita:2015ymn, Domenech:2023ryc, Alinea:2022ygr, Naruko:2019jzj, Heisenberg:2016eld, Papadopoulos:2017xxx, Bettoni:2013diz, Zumalacarregui:2013pma, Domenech:2015hka, Domenech:2015tca, Crisostomi:2016czh, BenAchour:2016cay, Domenech:2025eii, Gorji:2025ajb}. This leads us to natural question: \textit{Can inclusion of disformal couplings also remove the issues that we have previously found for the KR field?} In the following, we will investigate this question starting with the flat space-time. 

Let us first consider the $\beta$ coupling to the Ricci tensor, which is described in Subsection \ref{betaC}. By closely studying the interaction between the tensor modes and the KR field, we can notice that if we introduce the disformal couplings in the following way:
\begin{equation}\label{betadisformal}
    h_{\mu\nu}=\tilde{h}_{\mu\nu}-\frac{2\beta}{\Mpl^2}B_{\mu\alpha}B_{\nu}^{\;\;\alpha}. 
\end{equation}
Then, to the leading order, the Lagrangian density is given by:
\begin{equation}
    \begin{split}
        \mathcal{L}&=\frac{\Mpl^2}{8}\left(2h_{,\mu}^{\mu\nu}h^{\alpha}_{\nu,\alpha}-h_{\alpha\beta,\mu}h^{\alpha\beta,\mu}+2hh^{\mu\nu}_{,\mu\nu}+h_{,\mu}h^{,\mu}\right)-\frac{1}{12}\left(H_{\mu\nu\rho}H^{\mu\nu\rho}+3m^2B_{\mu\nu}B^{\mu\nu}\right)\\&
         -\frac{\beta}{2}B_{\mu\alpha}B^{\mu\alpha}\left(h_{,\gamma\delta}^{\gamma\delta}-\Box h\right) +\frac{\beta^2}{\Mpl^2}B_{\mu\alpha}B^{\mu\alpha}\Box\left(B_{\nu\beta}B^{\nu\beta}\right)\\&
        -\frac{\beta^2}{\Mpl^2} B_{\mu\gamma}B_{\nu}^{\;\gamma}\left[\partial_{\alpha}\partial^{\nu}\left(B^{\mu\delta}B^{\alpha}_{\;\delta}\right)+\frac{1}{2}\Box\left(B^{\mu\delta}B^{\nu}_{\;\delta}\right)-3\partial^{\mu}\partial^{\nu}\left(B^{\alpha\beta}B_{\alpha\beta}\right)\right]
    \end{split}
\end{equation}
where we have dropped the tilde on the metric perturbations. Therefore, we can see that by including the disformal coupling of the form (\ref{betadisformal}), the tensor mode contribution in the leading order is replaced with a coupling between the gravitational potentials and the KR field which coincides with the Ricci scalar coupling. In addition, one finds self-interactions between the KR field itself, appearing at the quartic order in perturbations. As we have previously seen, the leading order corrections that correspond to the KR field and the Ricci scalar do not give rise to the strong coupling of the tensor modes. Therefore, we will only have the strong coupling of the vector modes that appears at quartic order due to the perturbations in the KR field. Thus, the disformal couplings in this case remove the unnatural behavior in the tensor sector, and define a frame in which the matter should be minimally coupled to gravity -- \textit{the disformal frame}. 

It is easy to see that the Ricci tensor coupling of type $\sigma$ can be also cured by inclusion of following disformal couplings:
\begin{equation}\label{sigmadisformal}
    h_{\mu\nu}=\tilde{h}_{\mu\nu}-\frac{\sigma}{\Mpl^2}\left(\tilde{\varepsilon}_{\mu\beta\gamma\delta}B_{\nu}^{\;\;\beta}B^{\gamma\delta}+\tilde{\varepsilon}_{\nu\beta\gamma\delta}B_{\mu}^{\;\;\beta}B^{\gamma\delta}\right). 
\end{equation}
However, one can wonder if the same also applies to the couplings with the Riemann tensor. It turns out that in this case, the disformal couplings can be introduced only non-locally: 
\begin{equation}
h_{\mu\nu} = \tilde{h}_{\mu\nu} - \dfrac{2\gamma}{M_{pl}^2}\dfrac{\partial_\rho\partial_\sigma}{\Box}B_{\mu\rho}B_{\nu\sigma}, 
\end{equation}
for $\gamma$ type coupling, and 
\begin{equation}
h_{\mu\nu}=\tilde{h}_{\mu\nu}-\frac{4\lambda}{\Mpl^2}\frac{\partial_{\alpha\beta}}{\Box}\left(\tilde{B}_{\mu}^{\;\;\alpha}B_{\nu}^{\;\;\beta}+\tilde{B}_{\nu}^{\;\;\alpha}B_{\mu}^{\;\;\beta}\right)
\end{equation}
for the $\lambda$ type coupling, where the small brackets denote the symmetrization with respect to the indices $\beta$ and $\alpha$. Therefore, if one restricts only to local interactions, the only way to remove the strong coupling issue of the tensor modes for the Riemann tensor couplings is to set $\gamma$ and $\lambda$ to zero. 

Let us now discuss the background which is homogeneous and isotropic. In this case, we have found that the parity preserving couplings give rise to the runaway behavior of the vector modes. Let us focus on the coupling with the Ricci tensor $\beta$ since the corresponding disformal coupling for this case preserves locality. The generalization to the curved space-time is straight-forward: 
\begin{equation}
    g_{\mu\nu}=\bar{g}_{\mu\nu}-\frac{2\beta}{\Mpl^2}B_{\mu\gamma}B_{\nu}^{\;\;\gamma}
\end{equation}
where $\bar{g}_{\mu\nu}$ is the metric in the disformal frame. In this case, at the leading order the action becomes:
\begin{equation}
    \begin{split}
        S \simeq \int d^4x\sqrt{-\bar{g}}&\left\{\frac{\Mpl^2}{2}\bar{R}-\frac{1}{12}H_{\mu\nu\rho}H^{\mu\nu\rho}-\frac{m^2}{4}B_{\mu\nu}B^{\mu\nu}-(\alpha+\beta/2)\bar{R} B_{\mu\nu}B^{\mu\nu}-\zeta \bar{R}B_{\mu\nu}\Tilde{B}^{\mu\nu}\right.\\&\left.
        -\left[\gamma \bar{R}^{\mu\nu\rho\sigma}B_{\mu\nu}B_{\rho\sigma}+\lambda\Tilde{\bar{R}}^{\mu\nu\rho\sigma}B_{\mu\nu}B_{\rho\sigma}+\sigma \bar{R}^{\mu\nu}B_{\mu\lambda}\Tilde{B}_{\nu}^{\;\;\lambda}\right]\right\}
    \end{split}
\end{equation}

Therefore, we can notice that the inclusion of the disformal coupling for the $\beta$ interaction removes both strong coupling of the tensor modes and the runaway behavior that was originally related to the Ricci tensor coupling. It is important to stress that matter should be minimally coupled to gravity only after the disformal coupling is included to the theory, and defines the physical frame.   

At this point, one might wonder what happens to the parity-violating interaction, whose generalization of disformal coupling is given by: 
\begin{equation}
    g_{\mu\nu}=\bar{g}_{\mu\nu}-\frac{2\sigma}{\Mpl^2}\bar{\varepsilon}_{\mu\beta\gamma\delta}B_{\nu}^{\;\;\beta}B^{\gamma\delta}
\end{equation}
In this case the action becomes: 
\begin{equation}
    \begin{split}
        S \simeq \int d^4x\sqrt{-\bar{g}}&\left\{\frac{\Mpl^2}{2}\bar{R}-\frac{1}{12}H_{\mu\nu\rho}H^{\mu\nu\rho}-\frac{m^2}{4}B_{\mu\nu}B^{\mu\nu}-\alpha\bar{R} B_{\mu\nu}B^{\mu\nu}-(\zeta+\sigma/2) \bar{R}B_{\mu\nu}\Tilde{B}^{\mu\nu}\right.\\&\left.
        -\left[\beta \bar{R}^{\mu\nu}B_{\mu\alpha}B_{\nu}^{\;\;\alpha}+\gamma \bar{R}^{\mu\nu\rho\sigma}B_{\mu\nu}B_{\rho\sigma}+\lambda\Tilde{\bar{R}}^{\mu\nu\rho\sigma}B_{\mu\nu}B_{\rho\sigma}\right]\right\} \ .
    \end{split}
\end{equation}

Therefore, we find that the inclusion of the disformal coupling removes the runaway modes in the case of the coupling with the Ricci tensor. 

\section{Discussion} \label{section 6}

To date, Kalb-Ramond  fields have been of significant relevance to the theories of quantum gravity and cosmology that also potentially induce the parity violation of the gravitational waves. In this work, we have focused on their non-minimal coupling to gravity, considering the most general action that contains maximally two time-derivatives and which is quadratic in the KR field.  

Our analysis first explored the influence of non-linearities due to the non-minimal coupling in flat space-time. In this case, the KR field can be coupled to the Ricci scalar, and Ricci and Riemann tensors, as well as their parity-violating counterparts. However, we have reached an unexpected result: While all three types of the coupling give rise to the strong coupling of the vector components of the KR field, couplings with the Ricci and Riemann tensor in addition introduce the strong coupling of the tensor modes, regardless of the potential ghost-free combinations between the couplings in general backgrounds or inclusion of addition of higher-order self-interactions \cite{Heisenberg:2019akx, Horii:2025jen}. This means that no matter how small the mass of the KR is, the gravitational waves will be strongly coupled well below the Planck mass, yielding thus an unnatural result. By further studying the KR field in the homogeneous and isotropic background, we have then further found that the same type of coupling with parity preserving form also gives rise to the runaway modes for the vector modes, while the pseudoscalar mode at the same time acquires a tachyonic mass term.  In contrast to the parity-preserving couplings, parity-violating ones do not give rise to runaway modes or tachyonic instability.

It is useful to compare this to the case of a massive vector field. In contrast to the KR case, this field allows for two types of coupling if one restricts the analysis to the terms quadratic in the vector field with at most two derivatives -- the coupling with the Ricci tensor and the coupling to the Ricci scalar. It propagates the same number of modes as the KR field -- one longitudinal and two transverse modes. However, in the massless case, only the two transverse modes remain in the theory, whereas the KR field propagates a single pseudo-scalar mode. This has motivated an investigation of the possibility that the two theories might not be dual in \cite{Hell:2021wzm}, which was otherwise commonly stated in the literature (see eg. \cite{Kawai:1980qq, Smailagic:2001ch, Buchbinder:2008jf, Casini:2002jm, Dalmazi:2011df } ). In particular, the modes that are absent in the massless theory are expected to become strongly coupled and decouple from the remaining weakly coupled modes up to small corrections \cite{Hell:2025uoc}. Therefore, as noted in \cite{Hell:2021wzm}, in the presence of self-interactions, the notion of duality for the two theories breaks down due to the strong-coupling. We can see that this persists also for the non-minimal couplings. For the KR field, the perturbative approach breaks down for the vector modes and the tensor modes, while in the case of the {(massive)} vector field the scalar and tensor modes become strongly coupled. The contrast between the two theories also persists in the curved space-time. In the case of the KR field, two modes have a gradient instability and one (pseudo-scalar) acquires a tachyonic mass  due to the presence of the coupling with the (parity-preserving) Ricci and Riemann tensors. In the case of the vector field, the Ricci tensor makes only one scalar mode a runaway, while the two transverse modes have a tachyonic mass. 

Nevertheless, in \cite{Hell:2024xbv} it was shown that the inclusion of the disformal coupling can resolve both runaway and unnatural strong coupling for the tensor modes, if the external matter is coupled in the disformal frame. Therefore, despite the differences between the two theories, one might be inclined to investigate if the same type of disformal coupling can also resolve the issues appearing for the KR field. Surprisingly, we find that this is not fully the case. In particular, while this is possible for the coupling with the Ricci tensor,  the disformal couplings have to be non-local for the same trick to work also for the Riemann tensor. Therefore, it appears that one can either arrive to the non-local theory, or should set such couplings to zero.

\section*{Acknowledgments}

\textit{
    The authors would like to thank Tsutomu Kobayashi for discussions. This work was supported in part by JSPS KAKENHI No. {19K14702,}  ~24K00624, and 
by the World Premier International Research Center Initiative (WPI), MEXT, Japan.
}

\bibliographystyle{utphys}
\bibliography{paperbib}

\end{document}